\documentstyle[12pt,epsf,epsfig,rotate]{article}

\newcommand{\lan}{\langle}
\newcommand{\ran}{\rangle}

\renewcommand{\Re}{{\rm Re\,}}

\renewcommand{\d}{{\mbox d}}

\newcommand{ \be }{\begin{equation}}
\newcommand{ \ee }{\end{equation}}
\newcommand{\qEA}{q_{\rm EA}}

\begin{document}

\title{3D Spin Glass and 2D Ferromagnetic XY Model: a Comparison}

\author{David I\~niguez$^{(a)}$,
Enzo Marinari$^{(b)}$,\\ 
Giorgio Parisi$^{(c)}$ and Juan J. Ruiz-Lorenzo$^{(c)}$\\[0.5em]
$^{(a)}$  {\small  Departamento de F\'{\i}sica Te\'orica, 
Universidad de Zaragoza}\\
{\small   \ \  P. San Francisco s/n. 50009 Zaragoza (Spain)}\\[0.3em]
{\small   \tt david@sol.unizar.es}\\[0.5em]
$^{(b)}$  {\small  Dipartimento di Fisica and Infn, 
Universit\`a di Cagliari}\\
{\small   \ \  Via Ospedale 72, 07100 Cagliari (Italy)}\\[0.3em]
{\small   \tt marinari@ca.infn.it}\\[0.5em]
$^{(c)}$  {\small  Dipartimento di Fisica and Infn, Universit\`a di Roma}
   {\small {\em La Sapienza} }\\
{\small   \ \  P. A. Moro 2, 00185 Roma (Italy)}\\[0.3em]
{\small   \tt giorgio.parisi@roma1.infn.it  ruiz@chimera.roma1.infn.it}\\[0.5em]
}

\date{\today}

\maketitle

\begin{abstract}
We compare the probability distributions and Binder cumulants of the 
overlap in the 3D Ising spin glass with those of the magnetization in 
the ferromagnetic 2D XY model.  We analyze similarities and differences. 
Evidence for the existence of a phase transition in the spin glass 
model is obtained thanks to the crossing of the Binder cumulant. We 
show that the behavior of the XY model is fully compatible with the 
Kosterlitz-Thouless scenario. 
Finite size effects have to be dealt with by using great 
care in order to discern among two very different physical pictures 
that can look very similar if analyzed without large attention.
\end{abstract}  

\thispagestyle{empty}
\newpage

\section{\protect\label{S_INT}Introduction}

The issue of the existence of a phase transition in the three 
dimensional Ising spin glass has been a hard and difficult problem for 
more than two decades (see for example \cite{RIEGER,BOOK} and 
references therein).  Today there are clear numerical evidences 
favoring the existence of a low temperature broken phase 
\cite{KAWAYOUNG,MAPARURI,OUR,BOOK}, but a deeper understanding of the 
underlying physics still lacks.  It is clear, for example, that one is 
very close to the lower critical dimension (LCD), but understanding 
the details of the influence of such effect is highly non-trivial.

Determining for example the infinite volume limit of the 
Edward-Anderson order parameter \cite{MEPAVI} ($\qEA$) has been beyond 
reach till very recently, and the existence of the phase transition 
(both in $3$ and $4$ dimensions) has been established by exhibiting 
the crossing of the finite size Binder parameter.  One was able to 
show (for the $3D$ case see \cite{KAWAYOUNG,MAPARURI}) that curves of 
$g_{L}(T)$, $g_{L+1}(T)$ as a function of $T$ would cross at 
$T_{c}^{(L)}$, but it was impossible to determine the non-trivial 
limit of $g_{L}(T)$ for $L\to\infty$ at $T<T_{c}$ (and in the same way 
it was impossible to determine the large volume limit of $\qEA$).  
Only in the most recent period off-equilibrium techniques \cite{4DIM} 
and equilibrium simulations based on parallel tempering \cite{MARZUL} 
have allowed a statistically significant determination of the $4D$ 
infinite volume order parameter $\qEA$.

We start here by noticing that the behavior of the Binder parameter 
$g_{L}(q)$ in the $3D$ spin glass is very reminiscent of the one one 
finds in the $2D$ XY model (without quenched disorder), $g_{L}(m)$ 
(here $m$ is the magnetization).  Even for quite 
large lattices the curves for different lattice volumes are well 
split in the high $T$ phase, but seem to merge better than cross at 
low $T$.  Only on very large lattices one can exhibit a non-ambiguous 
(but always very small) crossing \cite{KAWAYOUNG,OUR}. 
The XY $2D$ model 
shows that the order parameter is zero in the thermodynamical limit 
only very slowly when increasing $L$.

The same kind of effect could be appearing in the $3D$ 
Edwards-Anderson spin glass, and in order to be sure one is dealing 
with a real phase transition with a non-zero order parameter one has 
to be very careful, and to show that is keeping under control possible 
contamination.  That is why we have decided to run a comprehensive 
comparison of the order parameter distributions for the $3D$ 
Edwards-Anderson spin glass and for the $2D$ XY model.  A detailed 
paper by Binder \cite{BINDER}, containing a study of the distribution 
functions for the Ising model, can be considered a methodological 
prototype to this kind of analysis, and can be used as a nice 
introduction to the finite size scaling techniques and ideas used in 
this setting.

Let us start by reminding the reader about some main points concerning 
the definition of the lower critical dimension.  The lower and the 
upper critical dimensions ($d_{l}$ and $d_{u}$ respectively) are 
important in qualifying a statistical system.  $d_{u}$ is the minimal 
dimension where mean field predictions hold (apart from logarithmic 
corrections), while the LCD, $d_{l}$, is the maximal dimension where the 
finite $T$ phase transition disappears.  A typical example is the 
usual Ising model, with $d_l=1$ and $d_u=4$ \cite{SFT}.

Since a $\phi^{3}$ term appears in the effective Hamiltonian of spin 
glasses (see \cite{MEPAVI,DKTREV} and references therein) one expects 
that the upper critical dimension is $d_u=6$.  One of the possible 
ways to determine the lower critical dimension is based on the 
determination of the critical exponent $\eta$.  One starts from the 
two points correlation function at the critical point, $T=T_{c}$, that 
for $|\vec{x}-\vec{y}|\to\infty$ behaves as

\be
  \langle\phi(\vec{x})\phi(\vec{y})\rangle \simeq 
  |\vec{x}-\vec{y}|^{-(d-2+\eta)}\ .
\ee
The LCD is defined by

\be
  d_{l}-2+\eta(d_{l})=0\ ,
  \protect\label{lower}
\ee
i.e. by the fact that there is no power law decay of the two point 
correlation function at the ($T=0$) critical point. A 
(replica-symmetric) $\epsilon$-expansion computation \cite{GREEN85} 
gives

\be
  \eta=-\frac{1}{3} \epsilon +1.2593 \epsilon^2 +2.5367 \epsilon^3\  ,
  \protect\label{eta}
\ee
where $\epsilon\equiv(6-d)$.  At order $\epsilon$ one is getting the 
promising estimate $d_{l}=3$, that collapses when including the higher 
order contributions, that do not allow any real solution for $d_{l}\le 
6$.  It is clear that because of one of the many reasons that could 
give troubles (for example replica symmetric breaking and poor 
convergence of the $\epsilon$-expansion) here the $\epsilon$-expansion 
is not helping in determining the LCD.

Equation (\ref{lower}) allows an estimate of $d_l$ based on numerical 
estimates of the $\eta$ exponent.  In four dimensions, with Gaussian 
couplings, one finds $\eta=-0.35\pm 0.05$ \cite{4DIM}, while in $3D$ 
$\eta=-0.40\pm 0.05$ \cite{OUR}.  The variation of $\eta$ with $d$ is 
small, and it seems safe to estimate $d_{l}\simeq 2.5$.  Even if this 
result is somehow peculiar (since in the field theoretical approach 
\cite{DKTREV} one does not see any trace of propagators with 
non-integer powers) it is confirmed by a mean field based analysis 
\cite{FRANZ} where one builds up an interface and looks at its 
behavior.  This mean-field computation gives $d_l=2.5$, in excellent 
agreement with the numerical estimate.

Numerical simulations in $3D$ \cite{KAWAYOUNG,OUR,BOOK} have now shown 
clearly that there is a finite $T$ phase transition, i.e.  that 
$d_{l}<3$.  The broken phase is mean field like, and understanding 
more details about it will be the goal of this paper.  It is also well 
established that in $2D$ one finds a $T=0$ phase transition (see 
\cite{BOOK} and references therein).  Summarizing, from state of the 
art numerical simulations one can deduce that $2\le d_l<3$.

Also the fact that $d_u=6$ is well supported by numerical results 
\cite{WAYO93,MEANFIELD}. In $6d$ one determines with good accuracy 
mean field exponents ($\gamma=1$, $\beta=1$ and $z=4$), with 
logarithmic corrections (that have been detected in the equilibrium 
simulations).

Here we will try to shed more light on the difficult numerical
simulations of the $3D$ Edwards Anderson spin glass.  The main problem
is probably in the fact that the system is very close to its LCD. So
the apparent merging of the Binder parameters in the low $T$ region,
that has only recently been disentangled to show a significant
crossing \cite{KAWAYOUNG,OUR}, is dramatically reminiscent of
the one one can observe in the case of a Kosterlitz-Thouless
transition.  We will try here to learn more about the effects of an
anomalous situation like the Kosterlitz Thouless (KT) one, by looking
for example to the Binder cumulant and to the overlap probability
distribution $P(q)$.  To do this we will discuss in same detail the
structure of the order parameter probability distribution in the $2D$
XY model without disorder.  We will stress how similar to the $3D$
spin glass things are at a first level of analysis, and where the
relevant differences can be found.  It is also remarkable that the
pure $2D$ XY model has a peculiar {\em aging} behavior \cite{CUKUPA}:
aging is one of the crucial features of spin glass systems, and its
qualification is of large importance.

In the next section we will define our models, the physical observable 
quantities, and we will give details about our numerical simulations.  
In section (\ref{S_RES}) we discuss our results, by following in 
parallel the $2D$ XY model without disorder and the $3D$ spin glass.  
In section (\ref{S_CONCLU}) we draw our conclusions.

\section{\protect\label{S_MOD}Models, Observables and Simulations}

We have studied the two dimensional XY model on a squared lattice.
The volume is denoted by  $V=L^2$, the hamiltonian is

\begin{equation}
  {\cal H}=-\sum_{<x,y>} \cos(\phi_x-\phi_y)\ ,
\end{equation}
where $<x,y>$ denotes a sum over nearest neighbor site pairs, $\phi$ 
is a continuous real variable, and periodic boundary conditions are 
imposed on the system.  This model shows an {\em infinite order} phase 
transition (with $\beta_c \approx 1.11$) \cite{GUPTA}, the 
Kosterlitz-Thouless transition.  In according with the 
Mermin-Wagner theorem \cite{SFT} there cannot be non-zero order 
parameters: the magnetization in the thermodynamical limit is zero for 
all $T>0$.  The KT transition is characterized by a change in the 
behavior of the two point correlation function, which goes from the 
exponential decay of the high temperature phase to the algebraic decay 
of the low temperature phase.  The whole low temperature phase 
($\beta> \beta_c$) is critical (the correlation length is infinite).

To simulate this model we have used the Wolff single cluster 
algorithm~\cite{WOLFF}.  The simulations have been run at five 
different values of $\beta$ in the low temperature phase: $\beta=1.3, 
1.4, 1.5, 1.7, 2.0$.  For each value of $\beta$ we have used the 
lattice sizes $L=8$, $16$, $32$, $64$, $128$, $256$.  For each value 
of $(\beta,L)$ we have used $200,000$ iterations of the single cluster 
algorithm, discarding the first half for thermalization.  The total 
CPU time required has been approximately one month on a 100 MHz 
Pentium based computer.

We have measured the probability distributions of

\begin{equation}
  m_1\equiv\frac{1}{V}\left| \Re \sum_x \exp(i \phi_x)\right| \ ,
\end{equation}
that we denote as $P_1(m_1)$.  We call $m_1^{\rm max}$ the value of 
$m_1$ where $P_1(m_1)$ is maximum and takes the value $P_1^{\rm 
max}\equiv {\rm Max}[P_1(m_1)]$.  We have looked in detail to the first 
and second moments of $P_1(m_1)$, $\langle m_1\rangle$ and $\langle 
m_1^2\rangle$.  We have also computed the Binder cumulant of the 
$P_1(m_1)$ distribution:

\begin{equation}
  B_1=\frac{1}{2}\left( 3-\frac{\langle m_1^4 \rangle}
  {\langle m_1^2 \rangle^2}  \right) \ .
\end{equation}
At low $T$ one can study the XY model by using the spin wave 
approximation, that neglects the role of vortices (since they are 
suppressed at low $T$). In the $T\rightarrow 0$ limit all the spins 
point in the same direction, and ($\theta$) is uniformly distributed, 
so that

\begin{equation}
  \langle m_1^p \rangle=\frac{1}{2\pi} \int_0^{2 \pi} \d\theta \cos^p
  \theta \ ,
\end{equation}
and the Binder cumulant at $T=0$ has the value

\begin{equation}
  B_1(T=0)=\frac{3}{4} \ .
\end{equation}

To determine the relevant scaling behavior
we use the fact that, for the XY model, $\chi \simeq L^{2-\eta(T)}$ and
$\langle m^2 \rangle\equiv\chi/L^2$, where 
in the spin wave approximation

\be
  \eta(T) = \frac{T}{2\pi}\ ,
  \protect\label{eq:eta}
\ee
is the anomalous dimension of the field. Moreover, since $P_1(m_1)$ is
a probability distribution, normalized to one, with non zero maximum
value ($m_1^{\rm max}$) (at least for finite values of the lattice sizes)  
and with $\lan m_1^2 \ran \simeq L^{-\eta}\to 0$ (at finite temperatures)
we have that\footnote{In the rest of the paper the symbol $A \simeq B$ means 
$A=O(B)$.}

\be
  P_1^{\rm max}\  m_1^{\rm max} \simeq 1 ,
\ee
independently of the lattice size, $L$. Since 

$$
m_1^{\rm max} \simeq \langle m_1 \rangle \simeq \langle m_1^2
\rangle^{1/2} \ ,
$$ 
we conclude that

\begin{eqnarray}
\nonumber
m_1^{\rm max} &\simeq& L^{-\eta(T)/2}\ ,\\ \nonumber
\langle m_1\rangle &\simeq& L^{-\eta(T)/2}\ ,\\ \nonumber
\langle m_1^2\rangle &\simeq& L^{-\eta(T)}\ ,\\ \label{theor}
P_1^{\rm max} &\simeq& L^{\eta(T)/2}\ .
\end{eqnarray}
The other model we have studied is the three-dimensional 
Ising spin glass with quenched random couplings $J$ distributed with a 
Gaussian law. The Hamiltonian is

\be 
{\cal H} \equiv -\sum_{<i,j>}\sigma_i J_{i,j} 
\sigma_j\ , 
\ee
where the spin are defined on a three-dimensional cubic lattice and
$<i,j>$ denotes 
a sum over nearest neighbor pairs.
 
As usual \cite{BOOK} we have simulated two real replicas ($\sigma$ and $\tau$) 
with the same quenched couplings, and we have measured the overlap

\be
  q(\sigma,\tau) \equiv \frac{1}{V}\sum_i \sigma_i \tau_i\ ,
\ee
and its probability distribution

\be
P(q)=\overline{\lan \delta(q-q(\sigma,\tau) \ran}
\ee
where, as usual, we denote the thermal average with
$\lan(\cdot\cdot\cdot)\ran$, and the average over the disorder distribution
with $\overline{(\cdot\cdot\cdot)}$. The Binder cumulant of the
probability distribution $P(q)$ is

\be
  g \equiv \frac{1}{2} 
  \left[
  3-\frac{\overline{\langle q^4\rangle}}{\overline{\langle 
  q^2\rangle}^2} \right] \ .
  \label{E-BINDER}
\ee
We have run $L=4$,$6$,$8$,$10$,$12$ and $16$ lattices with
$2048$, $2560$, $512$, $512$, $2048$ and $500$ samples respectively. We 
have used the supercomputer APE-100~\cite{APE}. 

For simulating the spin glass model we have used the simple tempering 
method for small lattices ($L\le 10$), and the parallel tempering 
scheme for large lattices ($L\ge 12$) (see \cite{TEMPERING,ENZO,BOOK} and 
references therein).  Thanks to that we have kept under control the 
level of thermalization reached by the system, that is in all cases 
very good (for a discussion of the standard criteria of control see 
\cite{ENZO}). We have checked that the equalities established 
numerically  in \cite{MAPARURI}, and proven by Guerra \cite{GUERRA} 
hold for our results, and that the $P(q)$ is well symmetric, 
supporting the reach of full thermalization.

\section{\protect\label{S_RES}Results}

\begin{figure}[htbp]
\begin{center}
\leavevmode
\epsfysize=250pt
\rotate[r]{\epsffile{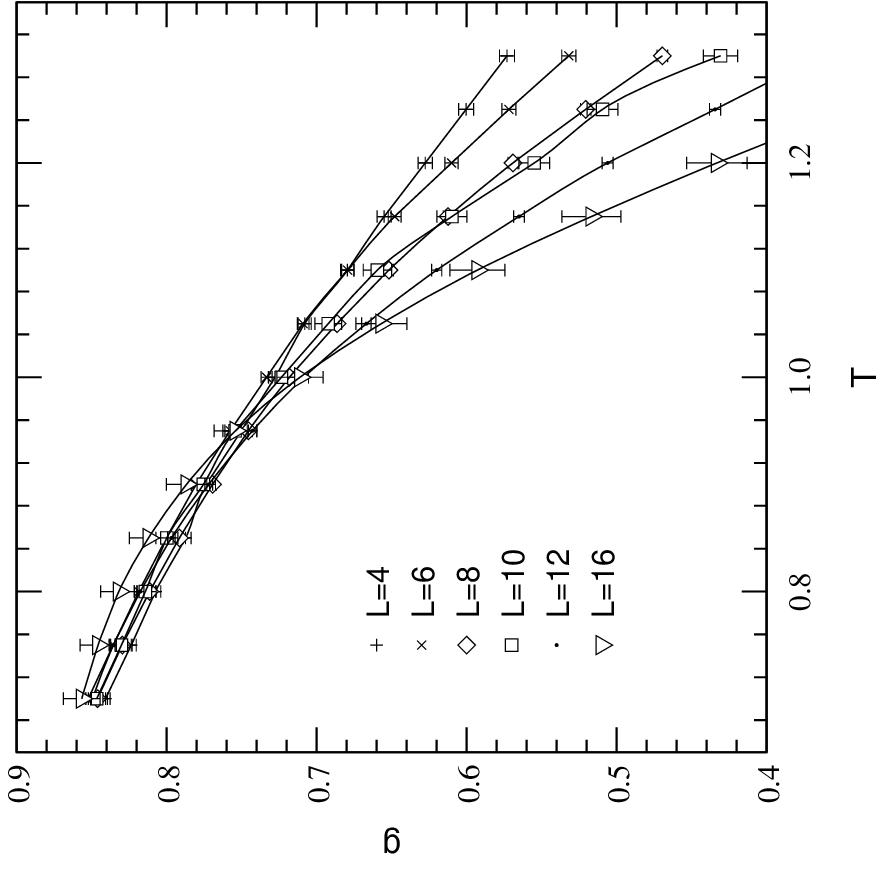}}
\end{center}
\caption[0]{Binder cumulant for the $3D$ Ising spin glass.
On the right, from top to bottom, curves and data points are for 
$L=4$, $6$, $8$, $10$, $12$ and $16$.}
\protect\label{fig:binder_sg}
\end{figure}

\begin{figure}[htbp]
\begin{center}
\leavevmode
\epsfysize=250pt
\centerline{\rotate[r]{\epsffile{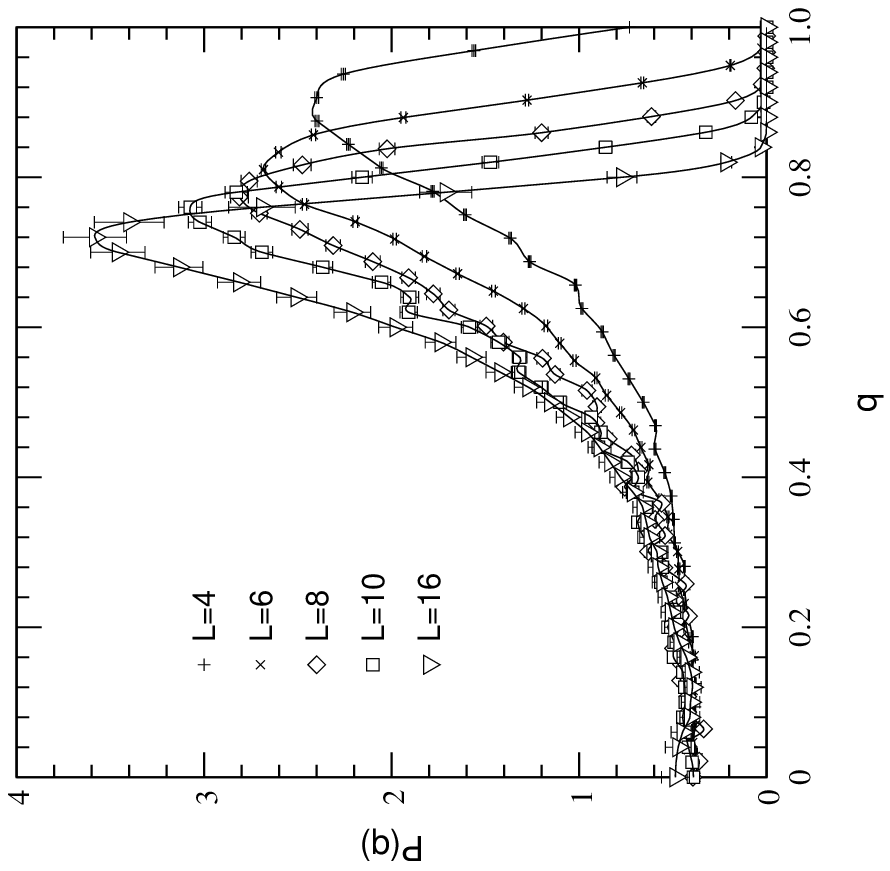}}}
\end{center}
\caption[0]{Probability distribution of the overlap, $P(q)$, for the $3D$
Ising spin glass. $T=0.7$
From right to left curves and data points are for
$L=4$,$6$,$8$,$10$ and $16$.}
\protect\label{fig:pq_sg}
\end{figure}

In figures (\ref{fig:binder_sg}) and (\ref{fig:pq_sg}) we show
the Binder cumulant and the probability distribution
of  the $3D$ Ising spin glass.

Let us discuss first the Binder cumulant.  In figure 
(\ref{fig:binder_sg}) there are two different regions.  In a high 
temperature region curves corresponding to different lattice 
sizes are clearly split (they tend to zero in the thermodynamical 
limit).  At small temperature (i.e.  for $\beta$ larger than 
$\beta_c^{\rm SG}\approx 1.0$), on small lattice sizes (up to $L=10$) 
curves coalesce, within our small error bars, in one.  It is 
interesting to notice that defining a Binder cumulant based on three 
different replicas \cite{INPARU} allows a somehow easier determination 
of the critical behavior.

Only when thermalizing a $L=16$ lattice (quite large for current 
standards, and impossible to thermalize deep in the critical region 
without the use of parallel tempering \cite{BOOK}) one is able to exhibit 
a clear crossing between, for example, the $L=8$ curve and the $L=16$
curve.  This implies the existence of a phase transition at
finite temperature with a non-zero order parameter, $\qEA\ne 0$ (see 
Kawashima and Young \cite{KAWAYOUNG}  
for the model with quenched 
binary couplings, $J=\pm 1$).

\begin{figure}[htbp]
\begin{center}
\leavevmode
\epsfysize=250pt
\centerline{\rotate[r]{\epsffile{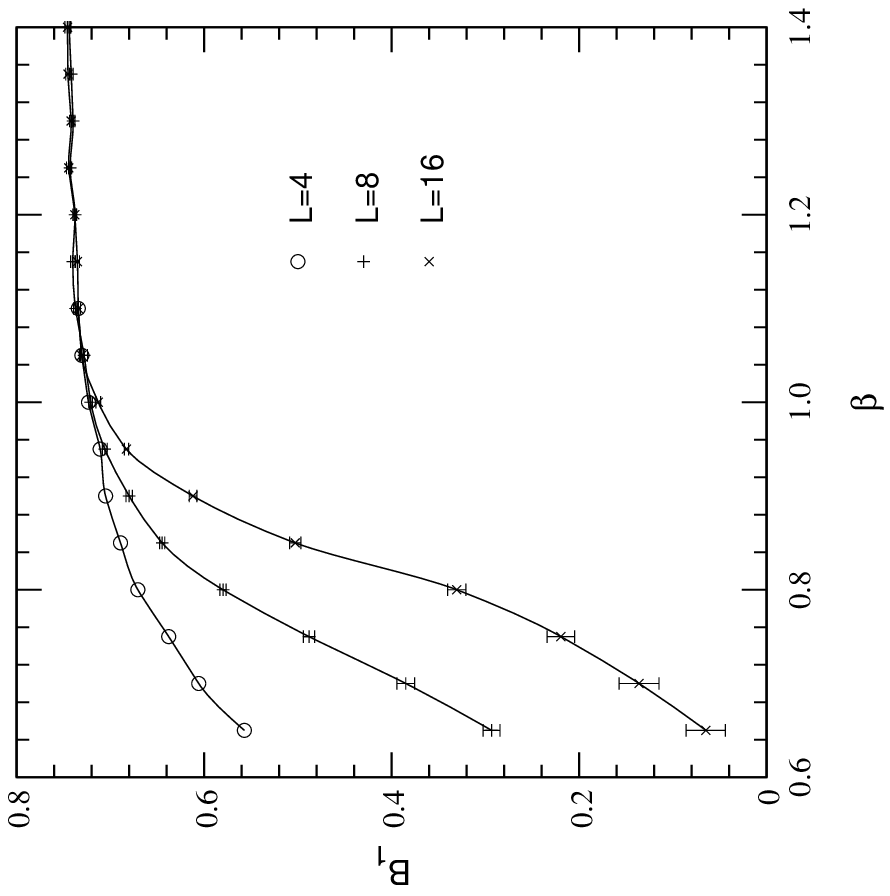}}}
\end{center}
\caption[0]{Binder cumulant, $B_1$, for the $2D$ XY model. From 
top to bottom, $L=4$, $8$ and $16$.}
\protect\label{fig:binder_xy}
\end{figure}

We can compare figure (\ref{fig:binder_sg}) with figure 
(\ref{fig:binder_xy}), where we show our numerical results for the 
Binder cumulant, $B_1$, for the $2D$ XY model.  Up to $L=10$ the XY 
model and the $3D$ Ising spin glass have a very similar behavior: 
again, within error bars, in the low temperature region all the curves 
for different lattice sizes collapse in a single curve, without any 
visible sign of finite size effects.

\begin{figure}[htbp]
\begin{center}
\leavevmode
\epsfysize=250pt
\centerline{\rotate[r]{\epsffile{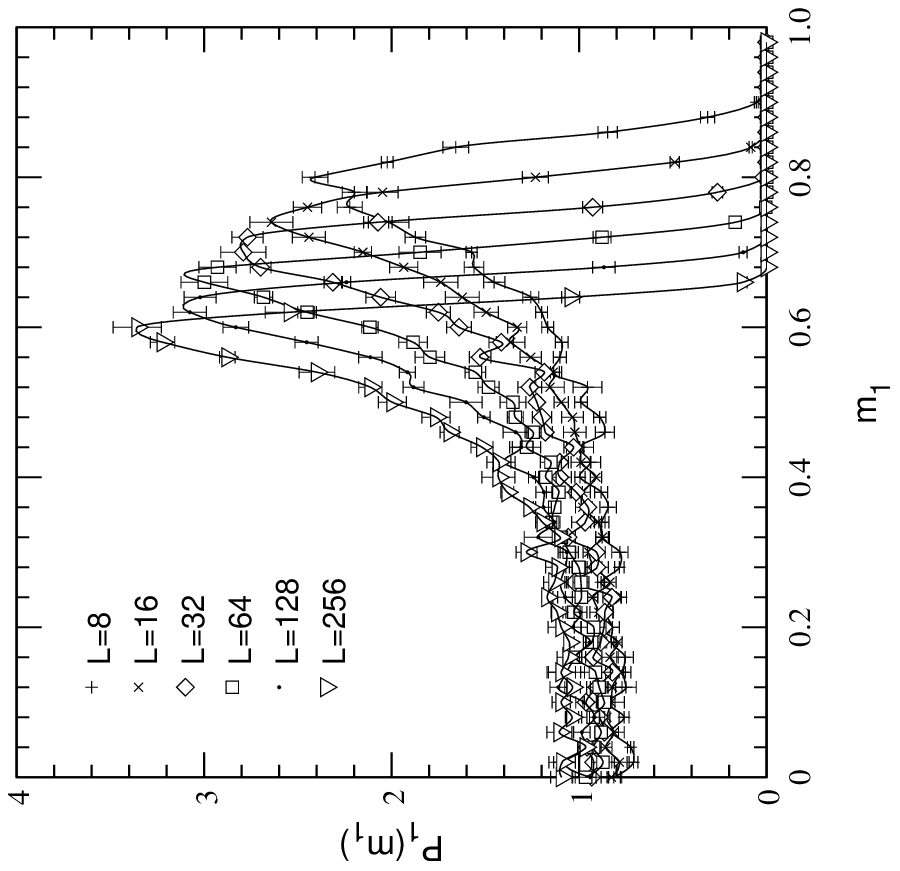}}}
\end{center}
\caption[0]{Probability distribution for the $2D$ XY model, $P_1(m_1)$,
 at $\beta=1.3$.}
\protect\label{fig:pd_xy_1}
\end{figure}

\begin{figure}[htbp]
\begin{center}
\leavevmode
\epsfysize=250pt
\centerline{\rotate[r]{\epsffile{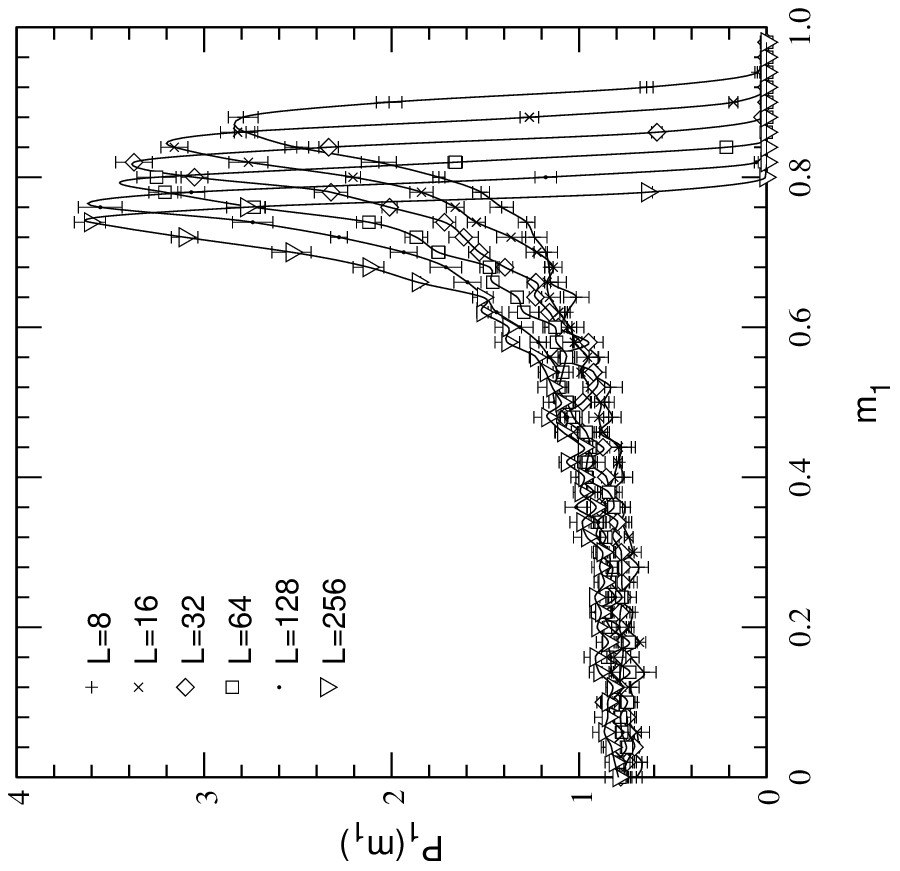}}}
\end{center}
\caption[0]{Probability distribution for the $2D$ XY model, $P_1(m_1)$,
 at $\beta=2.0$.}
\protect\label{fig:pd_xy_2}
\end{figure}

The behavior of the full probability distribution of the order 
parameter ($m_{1}$ for the $2D$ XY model and $q$ for the $3D$ Ising 
spin glass) is very similar. Figures (\ref{fig:pd_xy_1}) and 
(\ref{fig:pd_xy_2}), where we show $P_1(m_1)$ at $\beta=1.3$ and 
$\beta=2.0$ respectively, can be compared to the analogous figure for 
the spin glass $P(q)$, (\ref{fig:pq_sg}). The overall shapes are very 
similar. The peak shifts to the left, and in both cases $P(0)$ looks 
constant in our statistical precision.

We give in table (\ref{table:m_1}), at $\beta=1.3$, the expectation 
values of the observables shown figures (\ref{fig:pd_xy_1}) and 
(\ref{fig:pd_xy_2}).  By fitting these values by using a single power 
fit we find

\begin{eqnarray}                                    \nonumber
m_1^{\rm max}           &\simeq  & L^{-(0.08\pm 0.01)}\ ,\\ \nonumber
\langle m_1\rangle      &\simeq  & L^{-(0.09\pm 0.01)}\ ,\\ \nonumber
\langle m_1^2\rangle    &\simeq  & L^{-(0.17\pm 0.01)}\ ,\\ 
P_1^{\rm max}           &\simeq  & L^{+(0.09\pm 0.01)}\ .
\end{eqnarray} 
These results are in remarkable agreement among them and in good 
agreement with the spin wave
exact value $\eta(\beta=1.3)=0.12$.  
Corrections due to vortices are equivalent to a higher effective 
temperature~\cite{ITZDROUFF}, that undergoes here a $30\%$ shift.

We have also established that a power fit to a non-zero infinite 
volume order parameter of the form

\be
m_1^{\rm max}(L)=m_1^{\rm max}(\infty) + \frac{A}{L^B},
\ee
with $m_1^{\rm max}(\infty)$ different from zero and $A$ and $B$ 
constant is excluded by the data.

Figures (\ref{fig:pd_xy_1}) and (\ref{fig:pd_xy_2}) are interesting: 
they show a finite size non-trivial behavior that we know, from 
theoretical ideas (the Mermin-Wagner theorem), and from the analysis 
of the numerical data, will converge to a zero centered delta function 
limiting probability distribution in the infinite volume limit. 
This is the point we want to stress. Since the $3D$ spin glass has a 
very similar behavior (and even for the $4d$ model, where the crossing 
of the Binder cumulant is clear, it is non-trivial to show that 
$\qEA$ tends to a non-zero limit) it is crucial to understand where 
differences are.

We also want to stress that $P_1(m_1)$ shows a clear plateau, roughly 
$L$-independent, close to the $m_1\simeq 0$ region.  The plateau 
height grows with the lattice size (in a statistically significant way 
in our numerical data, see (\ref{table:m_1})).  This is one of the 
interesting results of this note: the infinite volume $2D$ XY $m_{1}$ 
delta function is constructed from the increasing finite volumes by a 
finite $m_{1}$ peak that shifts towards $m_1\simeq 0$, and by a 
plateau in the $m_1\simeq 0$ region that slowly increases with the 
lattice size, to eventually match the peak in the $m_1 = 0$ delta 
function.

\begin{figure}[htbp]
\begin{center}
\leavevmode
\epsfysize=250pt
\centerline{\rotate[r]{\epsffile{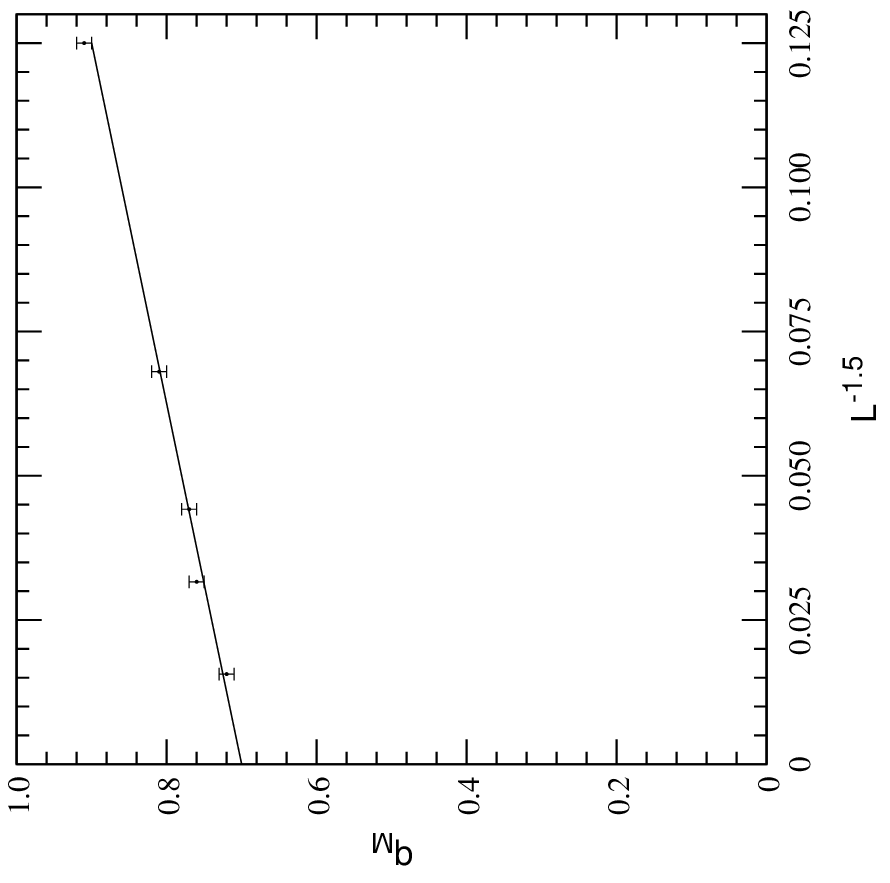}}}
\end{center}
\caption[0]{
Value of the overlap $q_M$ such that $P(q)$ is maximum  ($3D$ 
Ising spin glass).  The continuous line is the 
fit described in the text.  $T=0.7$.}
\protect\label{fig:q_sg}
\end{figure}

We have repeated this analysis for the overlap probability 
distribution $P(q)$ of the $3D$ spin glass (see figure 
(\ref{fig:pq_sg})).  The best scaling fit of the peak position, $q_M$, 
where the probability distribution is maximum, by a power law (the data 
are in table (\ref{table:q_m})) gives

\be
  q_M=(0.70\pm 0.02) +(1.6\pm 0.7) L^{-(1.5\pm 0.4)}\ ,
  \label{thermo}
\ee
where $T=0.7$.  In this fit we have used all lattice volumes ($L \le 
16$).  The fit had a $\chi^2/{\rm dof}=0.15$.  This thermodynamical 
value we get for $\qEA$ is close to the value that has been extracted 
from an off-equilibrium simulation ($q \simeq 0.7$) \cite{OUR}.  The 
best (two parameter) fit obtained by fixing $q_M=0.7$ (considered as 
an input from the dynamical simulations) gives compatible results with 
smaller errors.  In figure (\ref{fig:q_sg}) we show the $q_M$ data 
versus $L^{-1.5}$ (see also table (\ref{table:q_m})), and the curve from the 
best (two parameter) fit.

From the numerical data for the $3D$ spin glass (that are from a state 
of the art large scale numerical simulation) we cannot exclude the 
possibility of $\qEA=0$ in the infinite volume limit. We find that 
the best fit  (that uses in this case only $L \ge 6$ data) 

\be
  q_M=(1.0\pm 0.1) L^{-(0.12\pm 0.02)}
  \label{power}
\ee
is very good. So, even if the scenario of a non-zero overlap is 
favored (the static value is equal to the dynamic one, the exponent 
of a decay to $q=0$ is very small) in the $3D$ case we cannot use this 
limit to be sure of the existence of a phase transition with a 
non-zero order parameter (in $4d$ recent high statistics data allow to 
establish this evidence \cite{MARZUL}). The safe evidence for the 
existence of a phase transition in the $3D$ spin glass relies in this 
moment on the statistically significant crossing of the finite $L$ 
Binder cumulant \cite{KAWAYOUNG,OUR}, that makes visible fine details 
of the equilibrium probability distribution.

Also the behavior of $P(0)$ turns out to be the potential source of 
many ambiguities. We have seen that in the XY model it grows very 
slowly with the lattice size, in order to asymptotically contribute 
to the $m_{1}=0$ delta function. Also in the $3D$ spin glass, where 
in a mean-field like broken phase we expect a finite limit for $P(0)$ 
we observe a constant plateau with a (non necessarily statistically 
significant) growth for $L=16$. This behavior contributes to falsify 
the droplet model picture, where one would expect $P(0)$ to decrease 
with the lattice size.

\begin{table}
\centering
\begin{tabular}{|c||c|c|c|c|c|} \hline
$L$ &  $m_1^{\rm max}$  &$\langle m_1\rangle$ & $\langle m_1^2\rangle$
&   $P_1^{\rm max}$ & $P_1(0)$  \\ \hline \hline
16  &    0.74(1)& 0.485(6)  & 0.283(4) & 2.64(9) & 0.83(5)       \\ \hline
32  &    0.70(1)& 0.456(5)  & 0.260(3) & 2.79(8) & 0.93(2)    \\ \hline    
64  &    0.66(1)& 0.431(5)  & 0.231(3) & 3.00(9) & 0.97(6)    \\ \hline
128 &    0.62(1)& 0.405(4)  & 0.205(2) & 3.08(7) & 0.94(9)    \\ \hline
256 &    0.60(1)& 0.382(5)  & 0.183(3) & 3.36(11)& 1.09(6)     \\ \hline
\end{tabular}
\caption[0]{Numerical data for the 
$2D$ XY model, $\beta=1.3$. See the text for more details.}
\protect\label{table:m_1}
\end{table}

\begin{table}
\centering
\begin{tabular}{|c||c|c|} \hline
$L$ & $q_M$&   $P(0)$   \\ \hline \hline
4   &   0.91(1) &       0.398(3)\\ \hline
6   &   0.81(1) &       0.376(5)\\ \hline
8   &   0.77(1) &       0.39(2)\\ \hline
10  &   0.76(1) &       0.39(2)\\ \hline
16  &   0.72(1) &       0.49(7) \\ \hline       
\end{tabular}
\caption[0]{Numerical data for the 
$3D$ Ising spin glass, $T=0.7$.
See the text for more details.}
\protect\label{table:q_m}
\end{table}

\section{\protect\label{S_CONCLU}Conclusions}

We have shown that the two dimensional ferromagnetic XY model and the 
three dimensional Ising spin glass finite volume order parameter 
probability distributions behave very similarly.  The Binder cumulants 
on small lattice volumes show a similar {\em merging} at low $T$.  
Only on large lattices the $3D$ spin glass exhibits a crossing typical 
of a phase transition.  The results that we have discussed for the 
$XY$ model are completely compatible with the KT predictions.  We have 
analyzed the finite volume behavior of the peak of the finite volume 
order parameter probability distribution.  In the case of the XY model 
the preferred limit is zero.  In the spin glass case the preferred 
value is non-zero, and compatible with an off-equilibrium estimate, but 
from the present data one cannot rule out the possibility of the 
position of the peak going to zero in the infinite volume limit.

We have also established that $P(0)$ in the KT scenario has a finite 
volume non-zero value, that increases in the infinite volume limit.  
In finite volume one can then exhibit probability distributions with 
the same shape of that of a finite dimensional spin glass that have a 
trivial thermodynamic limit (a delta function in the origin).  
Analyzing finite size effects is crucial before reaching conclusions 
about the critical behavior.  In the $3D$ spin glass good evidence for 
the existence of a mean field like phase transition is based on a 
dynamical determination of the Edward-Anderson order parameter and on 
the crossing of the Binder cumulant on large lattices, but 
determining with good precision the shape of $P(q)$ on large lattice 
sizes will be important for making more details of the critical 
behavior crystal clear.

\section{\protect\label{S_ACKNOWLEDGES}Acknowledgments}

D. I\~niguez acknowledges an FPU grant from MEC.  
J. J. Ruiz-Lorenzo is supported by an EC HMC (ERBFMBICT950429) grant. 
We thank Paola Ranieri for many useful conversations.

\end{document}